\documentclass[superscriptaddress,aps,letterpaper,10pt,twocolumn,floats,showpacs,amsmath,amsfonts,amssymb,pre]{revtex4-2}

\usepackage{graphicx}
\usepackage[colorlinks=true,citecolor=blue]{hyperref}
\usepackage[caption=false]{subfig}
\usepackage{pstricks}
\usepackage{pst-node}
\usepackage{amsmath}
\usepackage{amsbsy}
\usepackage{verbatim}
\begin{document}


\title{Post-thermalization via information spreading in open quantum systems}

\author{Krzysztof Ptaszy\'{n}ski}
\affiliation{Institute of Molecular Physics, Polish Academy of Sciences, Mariana Smoluchowskiego 17, 60-179 Pozna\'{n}, Poland}
\email{krzysztof.ptaszynski@ifmpan.poznan.pl}
\author{Massimiliano Esposito}
\affiliation{Complex Systems and Statistical Mechanics, Department of Physics and Materials Science, University of Luxembourg, L-1511 Luxembourg, Luxembourg}

\date{\today}

\begin{abstract}
Thermalization in open systems coupled to macroscopic environments is usually analyzed from the perspective of relaxation of the reduced state of the system to the equilibrium state. Less emphasis is given to the change of the state of the bath. However, as previously shown for some specific models, during the thermalization the environment may undergo a nontrivial dynamics, indicated by the the change of its von Neumann entropy, at time scales even longer than the relaxation time of the system; here such a behavior is nicknamed as post-thermalization. We provide a more detailed analysis of this phenomenon by simulating the full dynamics of a variety of systems together with their environment. In particular, the post-thermalization is qualitatively explained as a result of reconversion of the initially built up correlation between the system and the bath into the correlation between the degrees of freedom in the environment. We also present exemplary systems in which such a reconversion is suppressed due to non-Markovian dynamics or the presence of interactions.
\end{abstract}

\maketitle

\section{Introduction} \label{sec:intr}
When a finite system is coupled to an extensive thermal environment it undergoes thermalization, namely, it relaxes to an equilibrium state determined by the macroscopic properties of the bath (such as its temperature or chemical potential). Starting from the 19th century much effort has been done to explain the emergence of the irreversible nature of such a process from the reversible, microscopic laws of classical and quantum mechanics. An important step in this direction has been done in Ref.~\cite{esposito2010} which explained the microscopic nature of the entropy production -- a basic quantity describing the thermodynamic irreversibility of the process -- by relating it to generation of correlations in the joint state of the system and the environment in result of their coupling. Specifically, the authors considered a joint unitary evolution of the system and the environment starting (in the moment $t=0$) from the initially uncorrelated state ${\rho_{SE}(0)=\rho_S(0) \otimes \rho_E^\text{eq}}$, where the density matrices $\rho_{S}(0)$ and $\rho_E^\text{eq}$ represent an arbitrary initial state of the system and the equilibrium state of the environment, respectively. The entropy production after a time $t$ was defined as
\begin{align} \label{entrprod}
\sigma \equiv \Delta S_S-\beta Q,
\end{align} 
where $\Delta S_S=S_S(t)-S_S(0)$ is the change of the von Neumann entropy of the system $S_S=-\text{Tr} (\rho_S \ln \rho_S)$, $\beta$ is the inverse temperature of the bath and $Q$ is the heat subtracted from the bath within the time interval $[0,t]$ (generalization to multiple-bath environments is straightforward). It was shown that this quantity can be, on the other hand, expressed as a sum of two information-theoretic terms~\cite{esposito2010, reeb2014}
\begin{align} \label{entrprodinf}
\sigma=I_{SE}+D(\rho_E||\rho_E^\text{eq}).
\end{align} 
The first contribution $I_{SE}=S_S(t)+S_E(t)-S_{SE}(t)$ is the mutual information describing the correlations built up between the system and the environment during their joint evolution. The second term $D(\rho_E||\rho_E^\text{eq})={\text{Tr} [\rho_E (\ln \rho_E-\ln \rho_E^\text{eq})]}$ is the relative entropy between the actual state of the environment $\rho_E \equiv \rho_E(t)$ and the initial equilibrium state $\rho_E^\text{eq}$. It describes the displacement of the environment from equilibrium. As shown in our recent study~\cite{ptaszynski2019}, for extended baths this displacement can be related to generation of correlations between the initially independent degrees of freedom within the environment. 

This leads to the question: which of these contributions is dominant? In the same work~\cite{ptaszynski2019} we pointed out that the system-environment mutual information is bounded from above by a corollary of the Araki-Lieb inequality~\cite{araki1970, jaeger2007}
\begin{align} \label{araki}
{I_{SE} \leq 2 \text{min} \{ S_S,S_E\}} \leq 2 \ln \dim(\hat{H}_S),
\end{align}
where $\dim(\hat{H}_S)$ is the dimension of the Hilbert space of the system. In contrast, the entropy production itself is unbounded and may well exceed the maximum value of $I_{SE}$, especially when the system is driven far from equilibrium. This, however, does not preclude that the mutual information $I_{SE}$ may be dominant in the opposite regime when the entropy production is smaller than or comparable to $2 \ln \text{dim} (\hat{H}_S)$. Such a case has been studied in Refs.~\cite{pucci2013, einsiedler2020}, where thermalization of a quantum Brownian particle described by the Caldeira-Leggett model has been investigated. In these studies an interesting phenomenon has been observed: whereas the system-environment mutual information $I_{SE}$ can be a dominant contribution to the entropy production at short times, it reaches a maximum at a certain point of time and later decreases to some residual value, being exceeded by the relative entropy $D(\rho_E||\rho_E^\text{eq})$. A similar observation has been done for a collisional model, where the system interacts repeatedly with different units of the bath~\cite{cusumano2018}. Interestingly, the decay of the system-environment mutual information, and the associated non-trivial dynamics of the environment, occurs at time scales longer than the relaxation time of the system; therefore, in this work we nickname this phenomenon as \textit{post-thermalization}. 

Here we provide a detailed analysis of this process. First, based on our previous study~\cite{ptaszynski2019}, we qualitatively explain the post-thermalization as a result of reconversion of the system-environment correlations into the correlations between the degrees of freedom in the environment. Such an interpretation has been already suggested for a collisional model in Ref.~\cite{cusumano2018}; we show that it is also valid for systems continuously coupled to a bath consisting of many degrees of freedom. A similar phenomenon of spreading (or scrambling) of initially local information across many degrees of freedom in many-body quantum systems, and its relation to thermalization, has recently received much attention~\cite{nandkishore2015, bohrdt2017, swingle2018, swan2019, hummel2019}. Second, we investigate whether the post-thermalization can be suppressed such that the system-environment mutual information remains a dominant contribution to the entropy production also at long times. It is shown that this can be indeed achieved either by the formation of the system-environment bound state or by the presence of non-quadratic interactions which leads to the emergence of non-trivial higher-order many-body correlations.

The paper is organized as follows. In Sec.~\ref{sec:entrcor} we remind the basics of the formalism relating the entropy production to the generation of correlations. In Sec.~\ref{sec:cormat} the correlation matrix approach, used to simulate the dynamics and the thermodynamics of noininteracting systems, is briefly reviewed. Section~\ref{sec:res} presents our original results, namely, the numerical simulations of the thermalization and the post-thermalization in selected open quantum systems. Finally, Sec.~\ref{sec:concl} brings conclusions following from our results. Appendix~\ref{sec:app} includes some additional remarks concerning the role of initial state of the system.

\section{Entropy production as correlation} \label{sec:entrcor}
To make the paper self-contained, let us first review the main results of Refs.~\cite{esposito2010, reeb2014, ptaszynski2019} relating the entropy production to the generation of correlations. Throughout the paper we take $k_B=\hbar=1$. We consider a generic open quantum system described by the Hamiltonian
\begin{align}
\hat{H}_{SE}=\hat{H}_{S}+\hat{H}_{E}+\hat{H}_{I},
\end{align}
where $\hat{H}_{S}$, $\hat{H}_{E}$ and $\hat{H}_{I}$ are the Hamiltonians of the system, environment and the interaction between them, respectively. The supersystem SE consisting of the system and the environment undergoes a unitary evolution given by the von Neumann equation 
\begin{align}
i \frac{d}{dt} \rho_{SE}=\left[ \hat{H}_{SE},\rho_{SE} \right].
\end{align}
As mentioned before, the evolution is assumed to start from the initially uncorrelated state ${\rho_{SE}(0)=\rho_S(0) \otimes \rho_E^\text{eq}}$. The initial state of the environment it taken to be a grand-canonical Gibbs state
\begin{align} \label{gibbs}
\rho_E^\text{eq} =Z_E^{-1} e^{-\beta \left(\hat{H}_E-\mu \hat{N}_E \right)},
\end{align}
where $\beta$ and $\mu$ are the inverse temperature and the chemical potential of the environment, respectively, $\hat{N}_E$ is the particle number operator of the environment and $Z_E=\text{Tr} \exp [-\beta (\hat{H}_E-\mu \hat{N}_E )]$ is the partition function. Here we focus on thermalization, and therefore consider an environment consisting of a single bath, but generalization to the case of multiple baths is straightforward~\cite{esposito2010}.  

The heat subtracted from the environment is defined as
\begin{align} \label{heat}
Q=-\text{Tr} \left[\hat{H}_E \left(\rho_{E}-\rho_{E}^\text{eq} \right) \right]+\mu \text{Tr} \left[ \hat{N}_E \left(\rho_{E}-\rho_{E}^\text{eq} \right) \right],
\end{align}
where the first contribution is the change of the energy of the bath (with a minus sign), and the second is the chemical work associated with a particle transfer. Using Eq.~\eqref{gibbs} one gets
\begin{align} \label{heatexp}
\beta Q &= \text{Tr} \left[\left(\rho_{E}-\rho_{E}^\text{eq} \right) \log \rho_{E}^\text{eq} \right] \\ \nonumber &=\text{Tr} \left(\rho_{E} \log \rho_{E}^\text{eq} \right)-\text{Tr} \left(\rho_{E} \log \rho_{E} \right) \\ \nonumber &+\text{Tr} \left(\rho_{E} \log \rho_{E} \right)-\text{Tr} \left(\rho_{E}^\text{eq} \log \rho_{E}^\text{eq} \right) \\ \nonumber &= -D(\rho_E||\rho_E^\text{eq})-\Delta S_E,
\end{align}
where $\Delta S_E=S_E(t)-S_E(0)$; here in the second step we just added and subtracted $\text{Tr} (\rho_{E} \log \rho_{E} )$. One may now insert the expression above to Eq.~\eqref{entrprod} to obtain Eq.~\eqref{entrprodinf}. Here we use the facts that the system and the environment are initially uncorrelated, and the unitary evolution does not change the von Neumann entropy of the supersystem $S_{SE}$, thus $\Delta S_S+\Delta S_E=I_{SE}$.

To get a further insight into the nature of the relative entropy contribution $D(\rho_E||\rho_E^\text{eq})$ let us assume that the bath is composed of $K$ independent degrees of freedom $k$ (e.g. spins or fermionic/bosonic energy levels), later referred to as modes:
\begin{align}
\hat{H}_E=\sum_k \hat{H}_k.
\end{align}
Then the initial state of the environment can be expressed as a product of Gibbs states of independent modes, 
\begin{align}
\rho_E^\text{eq} =\prod_{k} \rho_{k}^\text{eq},
\end{align}
where $\hat{N}_k$ is the particle number operator of the mode $k$ and
\begin{align}
\rho_{k}^\text{eq}=Z_k^{-1} e^{-\beta (\hat{H}_k-\mu \hat{N}_k)},
\end{align}
with $Z_k=\text{Tr} \exp[-\beta (\hat{H}_k-\mu \hat{N}_k)]$. Then, in analogy to Eq.~\eqref{heatexp}, we have
\begin{align} \label{heatexpk}
\beta Q &= \sum_k \text{Tr} \left[\left(\rho_{k}-\rho_{k}^\text{eq} \right) \log \rho_{k}^\text{eq} \right] \\ \nonumber \nonumber &= -\sum_k D(\rho_k||\rho_k^\text{eq})-\sum_k \Delta S_k,
\end{align}
where $\rho_k \equiv \rho_k(t)$ is the density matrix of the mode $k$ in the moment $t$. Comparing Eqs.~\eqref{heatexp} and~\eqref{heatexpk} the relative entropy term can be decomposed into a sum of two independently non-negative contributions~\cite{ptaszynski2019}
\begin{align} \label{reldec}
D(\rho_E||\rho_E^\text{eq})=D_\text{env}+I_\text{env}.
\end{align}
The first term,
\begin{align}
D_\text{env}=\sum_k D(\rho_{k}||\rho_{k}^\text{eq}) \geq 0,
\end{align}
measures the deviation of the modes from their equilibrium states. The second contribution,
\begin{align}
I_\text{env}=\sum_k S_k-S_E \geq 0,
\end{align}
is the multipartite mutual information between the modes. It measures the correlation built up between the degrees of freedom in the environment. As shown in our previous paper~\cite{ptaszynski2019}, for extended bath (i.e., in the thermodynamic limit) the contribution $D_\text{env}$ becomes negligible and the relative entropy $D(\rho_E||\rho_E^\text{eq})$ is dominated by the term $I_\text{env}$. In such a case the entropy production is clearly related to generation of the mutual information, either between the system and the environment or between the modes of the environment:
\begin{align}
\sigma \approx I_{SE}+I_\text{env}.
\end{align}
This illustrates the microscopic nature of the entropy production as a result of generation of correlations between the initially uncorrelated degrees of freedom.

\section{Correlation matrix method} \label{sec:cormat}
To evaluate the information-theoretic contributions to the entropy production such as $I_{SE}$ or $D(\rho_E||\rho_E^\text{eq})$ an exact characterization of the evolution of both the system and the environment is required. The computational complexity of such a problem increases exponentially with the size of the bath, and therefore the numerical studies are tractable only for relatively small environments with just a few or a dozen modes. Fortunately, for noninteracting fermionic systems much larger baths can be studied using the correlation matrix approach introduced by Peschel~\cite{peschel2003}. Let us here briefly review this formalism, focusing on a way to calculate the thermodynamic and information-theoretic quantities.

The correlation matrix approach is applicable to systems described by a quadratic Hamiltonian
\begin{align}\label{hamq}
\hat{H}_{SE}=\sum_{ij} \mathcal{H}^{SE}_{ji} c_i^\dagger c_j,
\end{align}
and initialized in the Gaussian state of the form~\cite{peschel2003}
\begin{align}
\rho_{SE}(0)=\mathcal{Z}^{-1} e^{-\sum_{ij} \mathcal{A}_{ij} c_i^\dagger c_j},
\end{align}
where $\mathcal{Z}=\text{Tr} [\exp(-\sum_{ij} \mathcal{A}_{ij} c_i^\dagger c_j)]$; here $c_i^\dagger$ and $c_i$ are the creation and the annihilation operators, respectively. In particular, the thermal state of the Hamiltonian~\eqref{hamq} is Gaussian. The unitary evolution generated by a quadratic Hamiltonian leaves the Gaussian character of the state unchanged. In such a case the state of the supersystem SE is at all times fully determined by the two-point correlation matrix $\mathcal{C}^{SE}$ with the elements~\cite{peschel2003}
\begin{align} \label{cormat}
\mathcal{C}^{SE}_{ij}=\text{Tr} \left(c_i^\dagger c_j \rho_{SE} \right).
\end{align}
This is a consequence of Wick's theorem stating that in noninteracting systems all higher-order correlations can be expressed by means of two-point correlations. Analogously, the reduced states of the system (environment) are given by submatrices $\mathcal{C}^S$ ($\mathcal{C}^E$) containing the elements $\mathcal{C}^{SE}_{ij}$ corresponding to the system (environment). Moreover, the evolution of the supersystem is described by a corollary of the the von Neumann equation~\cite{eisler2012}
\begin{align}
i \frac{d}{dt} \mathcal{C}^{SE}=\left[ \mathcal{H}^{SE}, \mathcal{C}^{SE} \right],
\end{align}
where $\mathcal{H}^{SE}$ is the matrix of the coefficients $\mathcal{H}^{SE}_{ij}$. Since the dimension of the correlation matrix $\mathcal{C}^{SE}$ increases only linearly with the size of the environment, the computational complexity of the problem is greatly reduced and numerical studies of baths with hundreds or thousands of modes are feasible~\cite{sharma2015, ptaszynski2019}.

Using the correlation matrix all the contributions to the entropy production can be calculated. For fermionic systems the von Neumann entropy $S_\alpha$ (${\alpha \in \{SE,S,E,k\})}$ is given by~\cite{sharma2015}
\begin{align} \label{vneuferm}
S_\alpha=-\sum_\sigma \left[ C^\alpha_\sigma \ln C^\alpha_\sigma+(1-C^\alpha_\sigma) \ln (1-C^\alpha_\sigma) \right],
\end{align}
where $C^\alpha_\sigma$ are eigenvalues of the matrix $\mathcal{C}^\alpha$. Using Eqs.~\eqref{heat}, \eqref{hamq} and~\eqref{cormat} heat can be evaluated as
\begin{align}
Q=-\text{Tr} \left[ \left(\mathcal{H}^E -\mu \right) \left(\mathcal{C}_E- \mathcal{C}_E^\text{eq} \right)\right],
\end{align}
where $\mathcal{H}^E$ is the submatrix of $\mathcal{H}^{SE}$ corresponding to the Hamiltonian $\hat{H}_E$; here we use $\langle \hat{N}_E \rangle = \text{Tr} (\hat{N}_E \rho_E ) =\text{Tr} (\mathcal{C}^E )$. Finally, though a direct analytic formula for the relative entropy $D(\rho_E||\rho_E^\text{eq})$ has been derived~\cite{colla2021}, it can be more easily evaluated using Eq.~\eqref{heatexp} as $D(\rho_E||\rho_E^\text{eq})=-\beta Q -\Delta S_E$.

While our paper focuses on fermionic systems, a similar formalism can be applied to the bosonic case; see Refs.~\cite{pucci2013, einsiedler2020, colla2021} for details.

\section{Results} \label{sec:res}
\subsection{Noninteracting resonant level (NRL)} \label{subsec:nrl}
\subsubsection{Post-thermalization}
Let us now present the original results concerning the behavior of the entropy production and its information-theoretic constituents during thermalization of some exemplary open quantum systems. We start our discussion with a single fermionic level coupled to a fermionic bath, referred to as the noninteracting resonant level model (NRL). The Hamiltonian of the system, environment and the interaction Hamiltonian read respectively as
\begin{align}
\hat{H}_S&=\epsilon_d c^\dagger_d c_d, \\
\hat{H}_E&=\sum_k \epsilon_k c^\dagger_k c_k, \\
\hat{H}_I&=\sum_{k} \left(\Omega c_d^\dagger c_{k} + \text{h.c.} \right),
\end{align}
where $\epsilon_i$ is the particle energy; here $\hat{H}_k=\epsilon_k c^\dagger_k c_k$. Energy levels of the bath $\epsilon_k$ are assumed to be uniformly distributed over the interval $[-W/2,W/2]$. The tunnel coupling $\Omega$ is parameterized by the coupling strength $\Gamma=2\pi \Omega^2/\Delta \epsilon$, where $\Delta \epsilon=W/(K-1)$ is the distance between the energy levels in the bath. The dynamics and the thermodynamics of the model are described using the correlation matrix method. The evolution of the correlation matrix is simulated iteratively using the equation
\begin{align}
\mathcal{C}^{SE}(t+\Delta t)=e^{-i \mathcal{H}^{SE} \Delta t} \mathcal{C}^{SE}(t) e^{i \mathcal{H}^{SE} \Delta t}.
\end{align}
We use the time interval $\Delta t=0.05/\Gamma$. The initial state is defined by the correlation matrix $\mathcal{C}^{SE}(0)={\text{diag} [\mathcal{C}^{SE}_{dd}(0),\mathcal{C}^{SE}_{11}(0),..,\mathcal{C}^{SE}_{KK}(0)]}$. The environmental levels are chosen to be initially thermalized: $\mathcal{C}^{SE}_{kk}(0)={f[\beta(\epsilon_k-\mu)]}$, where $f(x)=1/[1+\exp(x)]$ is the Fermi distribution function. The system is initialized in the empty state [$\mathcal{C}^{SE}_{dd}(0)=0$] and the coupling to the environment is switched on in the moment $t=0$; this can be interpreted as a quench of the system energy from $\infty$ to $\epsilon_d$. In our analysis we take $\epsilon_d=\mu$ to keep the entropy production small, i.e., smaller than the maximum value of $I_{SE}$ permitted by the Araki-Lieb inequality~\eqref{araki}.

%
\begin{figure}
	\centering
	\includegraphics[width=0.90\linewidth]{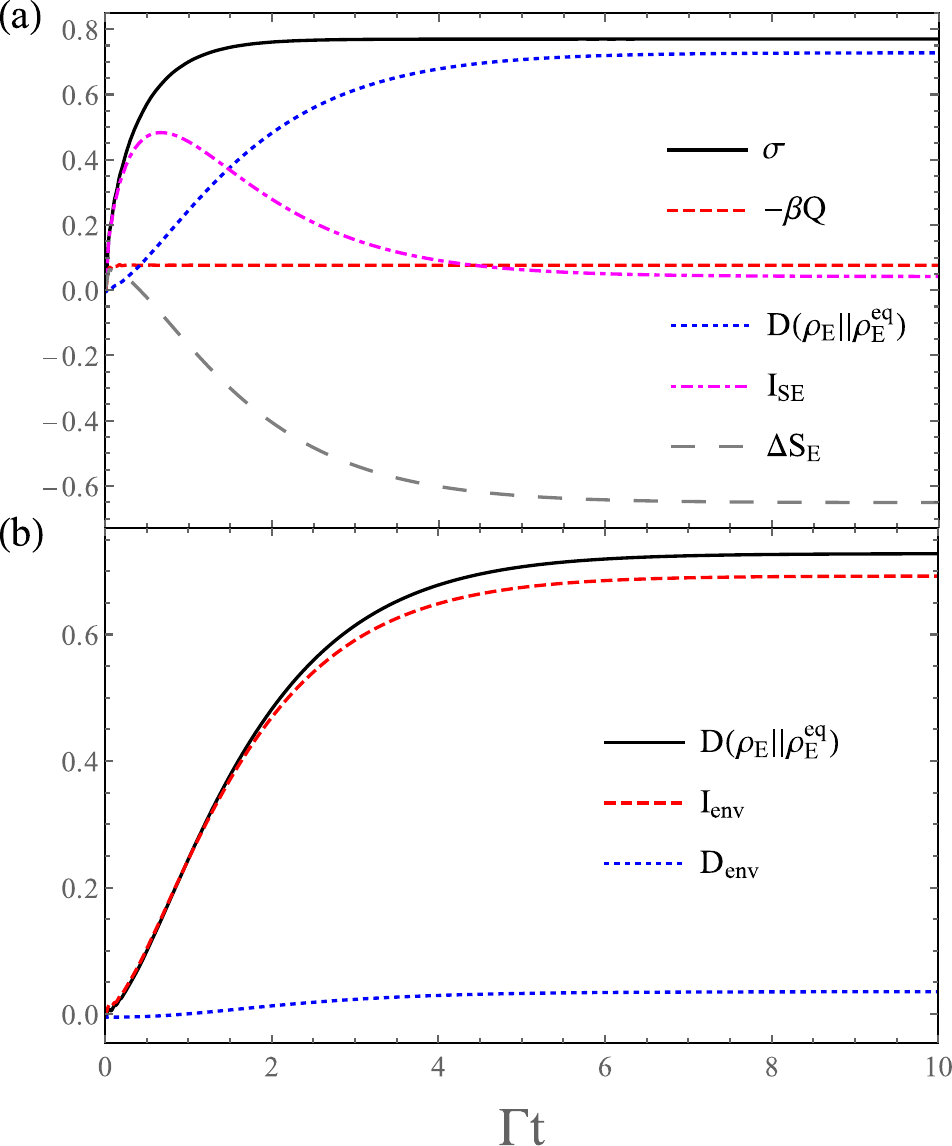}		
	\caption{The evolution of the entropy production and its constituents for NRL with the initially unoccupied system, ${\Gamma=0.1}$, ${\epsilon_d=\mu=0}$, ${\beta=1}$, ${W=20}$ and ${K=800}$.}
	\label{fig:singleres}
\end{figure}
%
The evolution of the entropy production and its constituents is presented in Fig.~\ref{fig:singleres}. Since we choose $\epsilon_d=\mu$ the small amount of heat is generated only due to switching on of the system-environment coupling, and the entropy production is dominated by the change of the von Neumann entropy of the system $\Delta S_S$. Analyzing the information-theoretic contributions to the entropy production one can clearly observe the phenomenon of post-thermalization described in the Introduction (Sec.~\ref{sec:intr}): whereas the system-environment mutual information $I_{SE}$ initially dominates, it later decreases to a residual value and is exceeded by the relative entropy $D(\rho_E||\rho_E^\text{eq})$. The decay of $I_{SE}$, growth of $D(\rho_E||\rho_E^\text{eq})$ and the corresponding decrease of the von Neumann entropy of the environment $S_E$ occurs at the time scale about three times longer than the thermalization time $1/\Gamma$. Fig.~\ref{fig:singleres}~(b) enables us to provide a qualitative interpretation of this phenomenon: the decay of $I_{SE}$ is associated with the reconversion of the system-environment mutual information into the mutual information between the environmental degrees of freedom $I_\text{env}$, which is a dominant contribution to $D(\rho_E||\rho_E^\text{eq})$.

\subsubsection{Dependence on the system-environment coupling}
%
\begin{figure}
	\centering
	\includegraphics[width=0.90\linewidth]{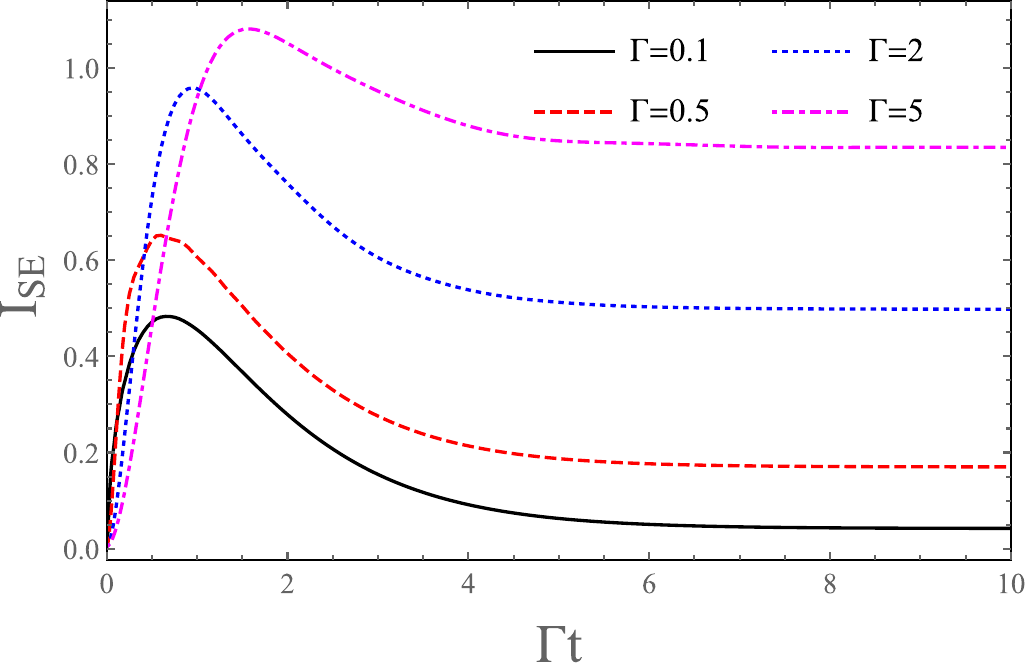}		
	\caption{The evolution of the system-environment mutual information $I_{SE}$ for different values of $\Gamma$. Other parameters as in Fig.~\ref{fig:singleres}.}
	\label{fig:singleresinfodg}
\end{figure}
%
%
\begin{figure}
	\centering
	\includegraphics[width=0.90\linewidth]{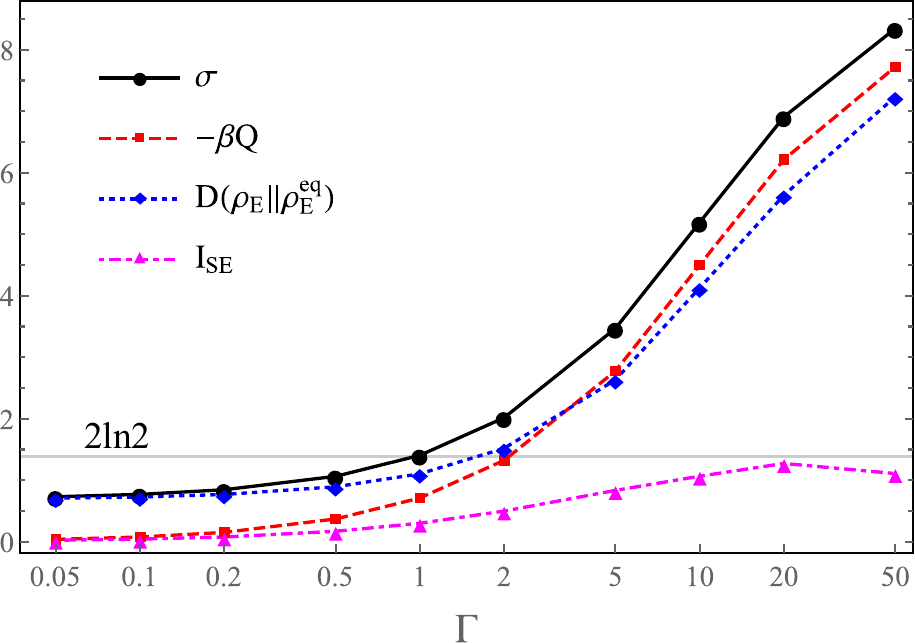}		
	\caption{The entropy production and its constituents for $\Gamma t=10$ as a function of $\Gamma$. Other parameters as in Fig.~\ref{fig:singleres}.}
	\label{fig:singleresodg}
\end{figure}
%
Let us now analyze the influence of the coupling strength to the environment $\Gamma$. As shown in Fig.~\ref{fig:singleresinfodg} the residual value of $I_{SE}$ increases with $\Gamma$. It is intuitive, since the equilibrium correlation between the system and the environment increases with the coupling strength. One may, therefore, be tempted to speculate that by increasing $\Gamma$ the residual value of $I_{SE}$ can be kept above the relative entropy $D(\rho_E||\rho_E^\text{eq})$. However, this is not the case since the amount of heat generated by the switching on of the coupling, and therefore the entropy production, also increases with $\Gamma$. This is illustrated by Fig.~\ref{fig:singleresodg} which shows the dependence of the long-time values of the entropy production and its constituents on the coupling strength. As one can see, although $I_{SE}$ may reach the Araki-Lieb bound for very large $\Gamma$, it is always exceeded by the the relative entropy $D(\rho_E||\rho_E^\text{eq})$

\subsubsection{Suppression of post-thermalization by the system-environment bound state} \label{sec:boundst}
%
\begin{figure}
	\centering
	\includegraphics[width=0.90\linewidth]{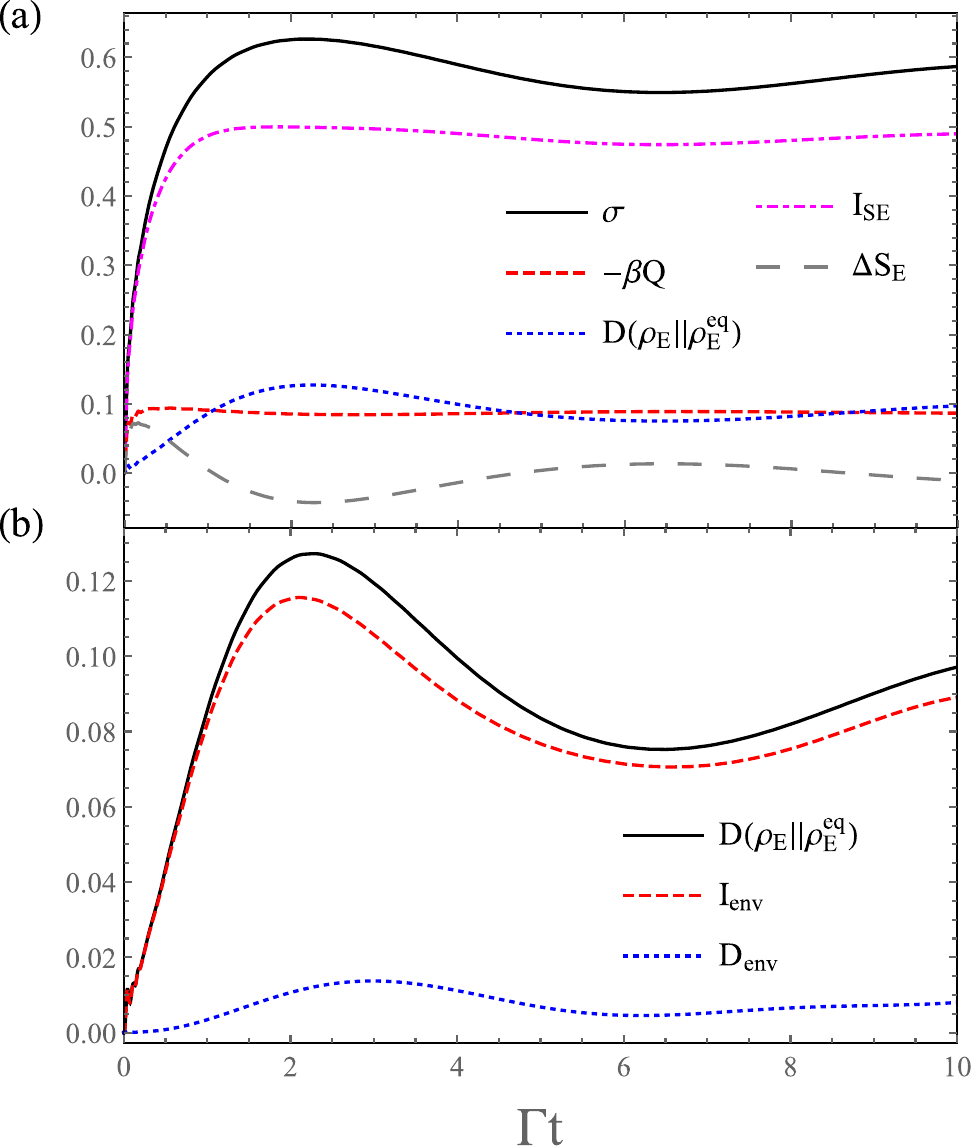}		
	\caption{The evolution of the entropy production and its constituents for NRL in the bound-state regime with the initially unoccupied system, $\delta=0$, ${\Gamma=0.1}$, ${\beta=1}$, ${W=10}$ and ${K=400}$.}
	\label{fig:singlereswid}
\end{figure}
%
%
\begin{figure}
	\centering
	\includegraphics[width=0.90\linewidth]{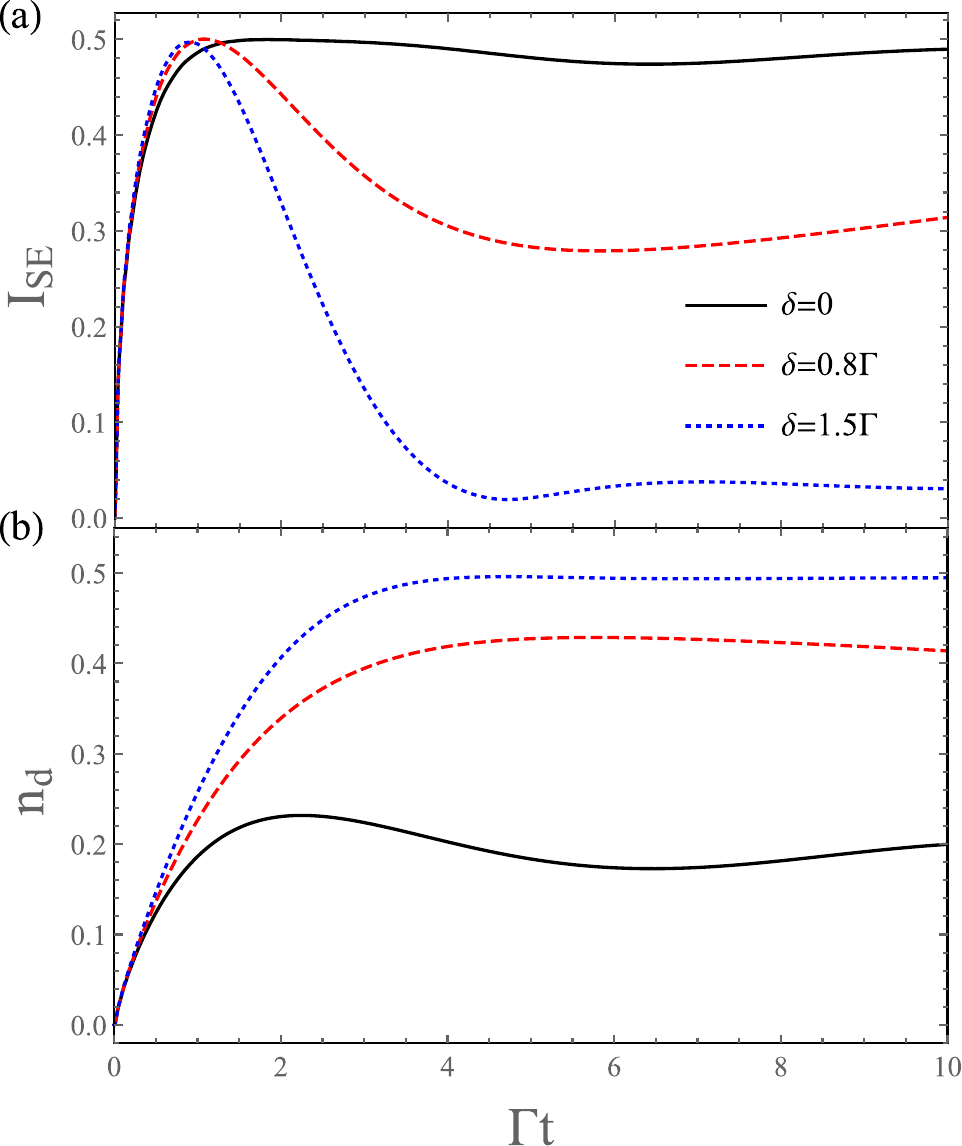}		
	\caption{The evolution of (a)~the system-environment mutual information $I_{SE}$ and (b) the occupancy of the system for different distances from the band edge $\delta$. Other parameters as in Fig.~\ref{fig:singlereswid}.}
	\label{fig:singlereswidinfoc}
\end{figure}
%
%
\begin{figure}
	\centering
	\includegraphics[width=0.90\linewidth]{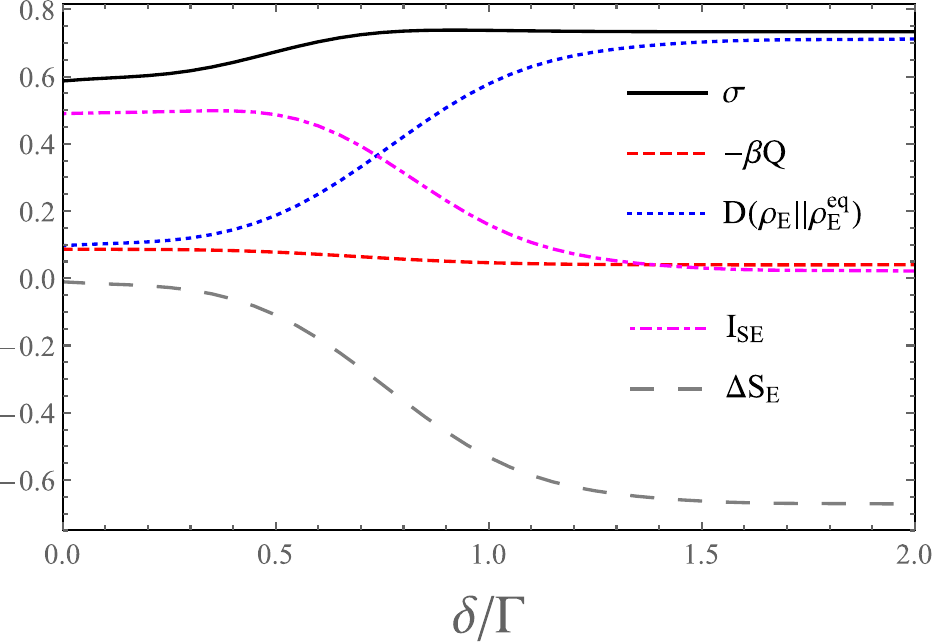}		
	\caption{The entropy production and its constituents for $\Gamma t=10$ as a function of $\delta$. Other parameters as in Fig.~\ref{fig:singlereswid}.}
	\label{fig:singlereswidodm}
\end{figure}
%
We will now demonstrate that the post-thermalization can be suppressed, and the system-environment mutual information can be kept above the relative entropy $D(\rho_E||\rho_E^\text{eq})$, in the strongly non-Markovian regime in which the system-environment bound state~\cite{cai2014, xiong2015, jussiau2019} is formed. This takes place when the energy level of the system $\epsilon_d$ is placed near a singularity of the density of states of the environment~\cite{cai2014}, e.g., near the band edge~\cite{jussiau2019}. Here we take $\epsilon_d=\mu=W/2-\delta$, where $\delta$ is the distance from the band edge. The evolution of the entropy production and its constituents for $\delta=0$ is presented in Fig.~\ref{fig:singlereswid}. As one can see, the decay of the system-environment mutual information is suppressed and $I_{SE}$ dominates over $D(\rho_E||\rho_E^\text{eq})$ also at long times. This is accompanied by a highly non-Markovian behaviour of the entropy production exhibiting temporal oscillations and temporary negative values of its time-derivative. Such oscillatory behavior, present even in the long time limit, is characteristic for the bound states~\cite{cai2014, xiong2015, alvaredo2020}.

The evolution of the system-environment mutual information for different values of $\delta$ is presented in Fig.~\ref{fig:singlereswidinfoc}. As shown, for large enough distance from the band edge one recovers the post-thermalization. Indeed, one may observe a transition from large to small residual values of $I_{SE}$ when $\delta$ is swept from $0.5\Gamma$ to $1.5\Gamma$ (Fig.~\ref{fig:singlereswidodm}). It is also worth noting that the formation of the bound state is accompanied by an imperfect thermalization of the system~\cite{cai2014, xiong2015}: the occupancy of the resonant level does not reach the equilibrium value $n_d=1/2$ [Fig.~\ref{fig:singlereswidinfoc}(b)].

\subsection{Interacting resonant level (IRL)} \label{subsec:irl}
Our previous discussion was concerned with a noninteracting system described by a quadratic Hamiltonian. Let us now investigate whether the presence of non-quadratic interactions changes the picture. To this goal we will analyze the interacting resonant level model (IRL) where, in comparison with NRL, the interaction Hamiltonian includes also a Coulomb interaction between the system and the bath:
\begin{align}
\hat{H}_I&=\sum_{k} \left(\Omega c_d^\dagger c_{k} + \text{h.c.} \right)+V \hat{n}_d \hat{N}_E,
\end{align}
where $V$ is the Coulomb coupling strength between the system and the environment and $\hat{n}_d$ is the particle number operator of the system. This system can be mapped on two other relevant open quantum systems, namely, the Ohmic spin-boson model and the anisotropic Kondo model~\cite{nghiem2016}.

Since the Hamiltonian of the model is not quadratic, the correlation matrix method is no longer applicable and the numerical simulation of the evolution of the full density matrix is required. Specifically, we use the iterative equation
\begin{align} \label{itvn}
\rho_{SE}(t+\Delta t) = e^{-i \hat{H}_{SE} \Delta t} \rho_{SE}(t) e^{i \hat{H}_{SE} \Delta t}.
\end{align}
In our simulation we take $\Delta t=0.2/\Gamma$. The necessity to consider the full density matrix severely limits the size of numerically tractable bath to just a few sites. However, as we will show, even in such a case a proper thermalization can be observed. To make the electrostatic energy independent of the size of the bath (for the initially empty system) we choose $\mu=0$ and parameterize the Coulomb coupling strength by the relation $V=U/(K/2-1)$. We also take $\epsilon_d=-U$ such that the system relaxes to the state with occupancy $n_d=1/2$ independently of $U$. Furthermore, due the small size of the bath a relatively large coupling strength $\Gamma$ is taken such that the system thermalizes before the Poincar\'{e} recurrence time is reached, for which the exponential relaxation breaks down and the system tends to return to the initial state~\cite{pucci2013}. Additionally, a relatively small bandwidth $W$ is required such that the bath level separation is small enough to resemble the continuous density of states of the environment.

%
\begin{figure}
	\centering
	\includegraphics[width=0.90\linewidth]{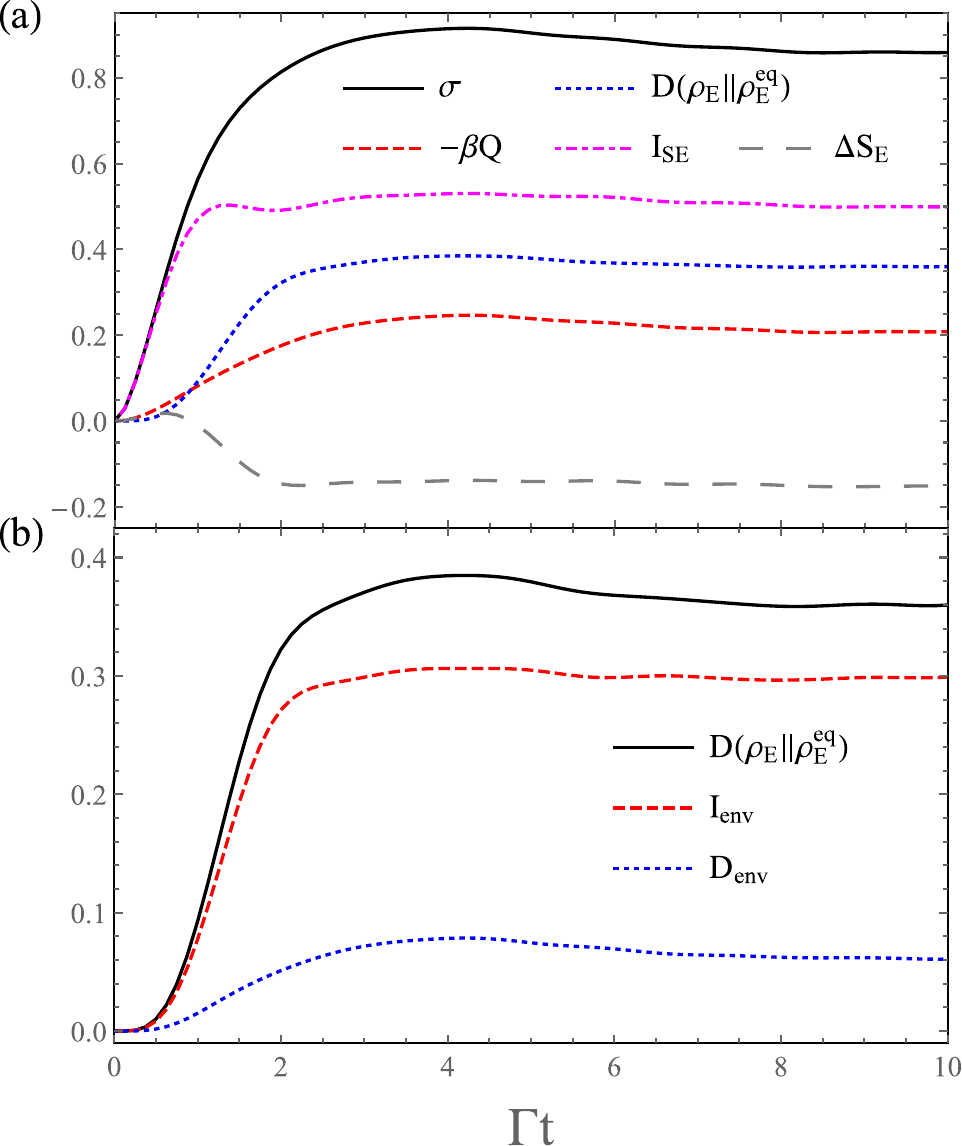}		
	\caption{The evolution of the entropy production and its constituents for IRL with the initially unoccupied system, ${\Gamma=1}$, ${U=3}$, ${\epsilon_d=-U}$, ${\mu=0}$, ${\beta=1}$, ${W=3}$ and ${K=7}$.}
	\label{fig:irl}
\end{figure}
%
The evolution of the entropy production and its constituents for a relatively large $U=3$ is presented in Fig.~\ref{fig:irl}. As shown, although some deviations from the exponential relaxation are observed, the behavior of the entropy production is relatively similar to the observed for NRL. The amount of generated heat is here larger due to the increased coupling strength. Most importantly, however, the decay of the system-environment mutual information is no longer observed and it dominates the relative entropy $D(\rho_E||\rho_E^\text{eq})$ also at long times. 

%
\begin{figure}
	\centering
	\includegraphics[width=0.90\linewidth]{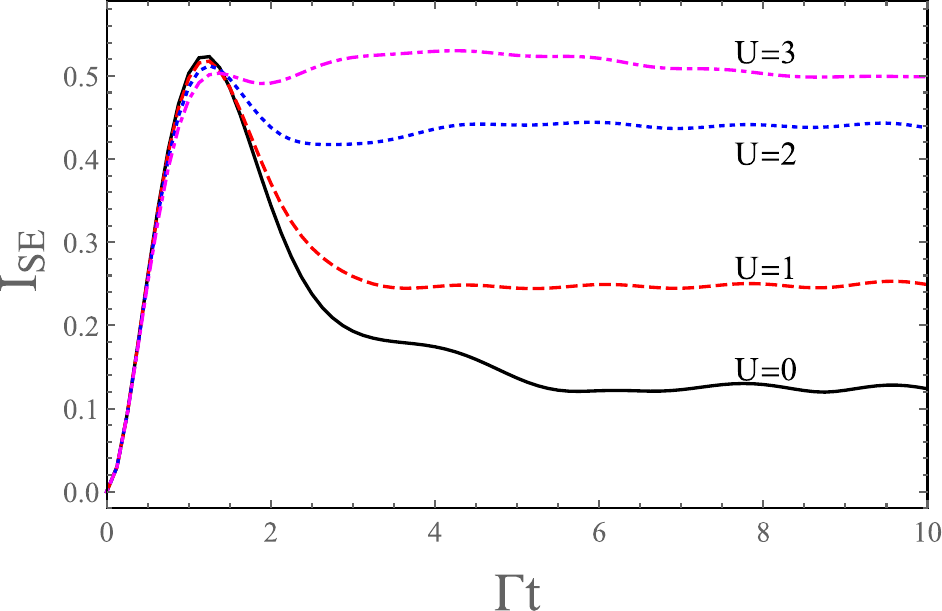}		
	\caption{The evolution of the system-environment mutual information $I_{SE}$ for different values of $U$. Other parameters as in Fig.~\ref{fig:irl}.}
	\label{fig:irlinfodu}
\end{figure}
%

%
\begin{figure}
	\centering
	\includegraphics[width=0.90\linewidth]{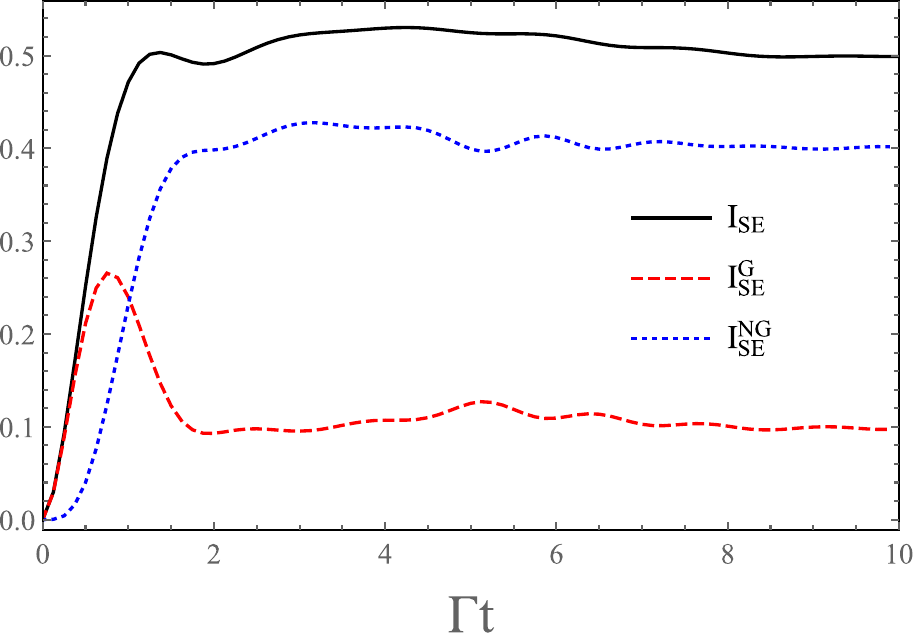}		
	\caption{The evolution of the Gaussian and the non-Gaussian contribution to the system-environment mutual information. Parameters as in Fig.~\ref{fig:irl}.}
	\label{fig:irlinfng}
\end{figure}
%
The dependence of the evolution of the system-environment mutual information on the Coulomb coupling is presented in Fig.~\ref{fig:irlinfodu}. As one can see, the residual value of $I_{SE}$ increases with $U$. For the noninteracting system the post-thermalization is clearly observed. This demonstrates that the suppression of decay of $I_{SE}$ for the interacting system is not a result of the finite size of the bath, small bandwidth or large $\Gamma$ but rather a consequence of the Coulomb coupling itself. To explain the origin of this phenomenon let us note that the presence of Coulomb interaction makes the system-environment state non-Gaussian: it is no longer fully characterized by a two-point correlation matrix, but rather the higher-order many body correlations are also important. The degree of non-Gaussianity can be characterized using the concept of the Gaussian von Neumann entropy $S^G_\alpha$ defined as the von Neumann entropy calculated with Eq.~\eqref{vneuferm} for a correlation matrix of a generic (possibly non-Gaussian) state. It was shown that $S^G_\alpha - S_\alpha = D(\rho_\alpha||\rho^G_\alpha) \geq 0$, where $\rho^G_\alpha$ is the reference Gaussian state, i.e., the Gaussian state with a same correlation matrix as $\rho_\alpha$~\cite{genoni2008}; the relative entropy $D(\rho_\alpha||\rho^G_\alpha)$ is commonly used to quantify the non-Gaussianity of the state. Let us also decompose the system-environment mutual information into the Gaussian and the non-Gaussian contributions $I_{SE}^G$ and $I_{SE}^{NG}$,
\begin{align}
I_{SE}=I_{SE}^G+I_{SE}^{NG},
\end{align}
where $I_{SE}^G=S_S^G+S_E^G-S_{SE}^G$. The evolution of these contributions for $U=3$ is presented in Fig.~\ref{fig:irlinfng}. As shown, the Gaussian contribution $I_{SE}^G$ follows the same behavior as the system-environment mutual information for the noninteracting case: it first increases, but later decays to some residual value. However, this is compensated by the increase of the non-Gaussian contribution $I_{SE}^{NG}$ (absent in the noninteracting system) such that a relatively high value of $I_{SE}$ is preserved. This qualitatively explain the difference with NRL: whereas two-point system-environment correlations eventually decay in both cases, for IRL the correlation between the system and the environment is preserved in the form of higher-order many body correlations.

Let us here make a brief comment about the potential role of initial conditions. Since in interacting systems non-Gaussian correlations are responsible for the preservation of the system-environment mutual information, one may wonder whether the presence of initial non-Gaussian correlations may influence the post-thermalization even in the noninteracting case; to be precise, the initial non-Gaussianity may be present only in the system, since the thermal state of a non-interacting environment has to be Gaussian. As shown in the Appendix~\ref{sec:app}, however, even for initially non-Gaussian states the post-thermalization is still observed in noninteracting systems. It therefore appears that interactions not only lead to generation of the higher-order system-environment correlations, but are also necessary for their preservation.

\subsection{Central spin model} \label{subsec:censpin}
Finally, let us show that the decay of the system-environment mutual information is strongly suppressed in the model of a single coupled to the spin bath (the central spin model); we focus on spins 1/2 for the sake of simplicity. It is described by the Hamiltonian 
\begin{align}
\hat{H}_S&=h_d \hat{s}_d^z, \\
\hat{H}_E &= \sum_k h_k \hat{s}_k^z, \\
\hat{H}_I&=\sum_{k} \left(J \hat{s}_d^+ \hat{s}_{k}^- + \text{h.c.} \right),
\end{align}
where $h_j$ is the value of the magnetic field acting on the spin $j$, $\hat{s}_j^z$ is the spin z-operator acting on the spin $j$ and $\hat{s}_j^\pm=\hat{s}_j^x \pm i \hat{s}_j^y$ are the spin raising ($\hat{s}_j^+$) and lowering ($\hat{s}_j^-$) operators. To make the structure of the system-environment Hamiltonian analogous to the previously analyzed fermionic models, the local magnetic fields in the bath $h_k$ are taken to be uniformly distributed over the interval $[-W/2,W/2]$ and the exchange coupling is parameterized by the coupling strength $\Gamma=2\pi J^2/\Delta h$ with $\Delta h=W/(K-1)$. Furthermore, to keep the amount of the generated heat small we take $h_d=0$. The bath is initialized in the canonical state, which corresponds to taking $\mu=0$ in Eq.~\eqref{gibbs}. As in the previous case, the dynamics of the system is described by Eq.~\eqref{itvn} with $\Delta t=0.2/\Gamma$.

%
\begin{figure}
	\centering
	\includegraphics[width=0.90\linewidth]{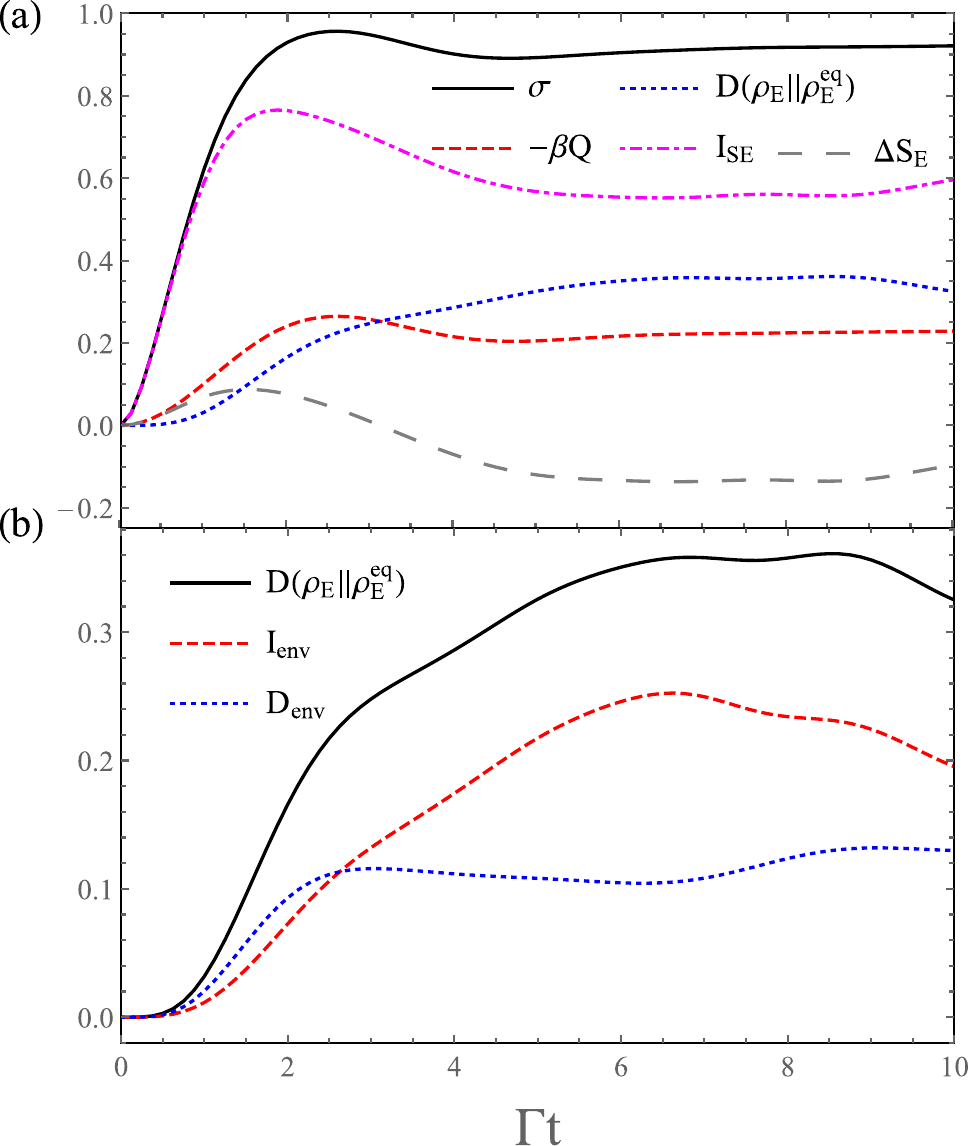}		
	\caption{The evolution of the entropy production and its constituents for the central spin model with the system initialized in the $|\! \! \downarrow \rangle$ state, ${\Gamma=1}$, ${h_d=0}$, ${\beta=1}$, ${W=3}$ and ${K=7}$.}
	\label{fig:spin}
\end{figure}
%
As shown in Fig.~\ref{fig:spin}, similarly to the interacting resonant level model, the decay of the system-environment mutual information is suppressed and it exceeds the relative entropy $D(\rho_E||\rho_E^\text{eq})$ also at long times. This can be explained by employing the correspondence between spins and fermions. Using Jordan-Wigner transformation, the considered setup -- with a system coupled to each spin of the bath -- can be mapped onto an interacting fermionic Kondo model~\cite{backens2019}. As shown in the previous subsection, interactions lead to generation of nontrivial higher-order correlations which are responsible for the preservation of the system-environment mutual information. 
	
Finally, let us note that a mapping to noninteracting fermions is, on the other hand, provided for one-dimensional spin chains with Ising or XY interactions, where each spin is coupled to at most two neighbors. In such a case (with a small part of the chain defined as the system and the rest taken as the bath) the post-thermalization can be still observed. This implies that information dynamics of spin models may be strongly affected by the topology of the underlying Hamiltonian. Interestingly, however, the presence of post-thermalization in spin chain models may also depend on the initial state of the system. Since this issue goes beyond the main focus of the paper, it is discussed in more detail in the Appendix~\ref{sec:app}.

\section{Conclusions} \label{sec:concl}
It is now well established that the entropy production is related to generation of correlations between the initially independent degrees of freedom in the system and the environment. This study demonstrates that the nontrivial processes of information conversion, changing the microscopic character of the entropy production, can occur even at time scales longer than the thermalization time, when the entropy production, macroscopically defined by Eq.~\eqref{entrprod}, is already well saturated. The evolution of the microscopic form of the entropy production may be related to the reconversion of the system-environment correlations into the intra-environment correlations. This process may be suppressed by the the non-Markovian dynamics, presence of non-quadratic interactions, or for specific types of the bath, such as spin baths.

Apart from deepening our understanding of the microscopic nature of the entropy production, our work may be relevant for the study of dynamics of closed quantum many-body systems, e.g., thermalization after the quench. Many studies focus on the prethermalization~\cite{gring2012, mori2018}, i.e., relaxation to a quasi-stationary nonequilibrium state before reaching the thermal equilibrium. Our study points out, instead, that even when the system seams to be already well thermalized, some non-trivial microscopic processes may occur, changing the character of many-body correlations. Indeed, the entanglement growth taking place after the local equilibration has been previously observed for quenched holographic systems~\cite{liu2014}. Such processes may potentially provide additional information about the intra-system interactions. Though our study focuses on the mutual information, it may be worth investigating whether the post-thermalization can be revealed by experimentally measurable quantities, e.g., out-of-time-order correlators~\cite{swingle2018, swan2019}.

Another topic worth exploring may be the connection between the behavior of the system-environment mutual information and the Markovian or the non-Markovian character of the reduced dynamics~\cite{breuer2002}. There is not necessarily a contradiction between the observed long time scale of decay of $I_{SE}$ and the validity of the Born-Markov approximation used when deriving master equations~\cite{breuer2002} because the latter only uses the assumption of weak correlations to predict the system evolution. Nevertheless, the results of Sec.~\ref{sec:boundst} suggest that the non-Markovian dynamics may be associated with the suppression of decay of the system-environment mutual information. Whether this conclusion can be generalized beyond the specific case analyzed is an open question.

\begin{acknowledgments}
K. P. is supported by the National Science Centre, Poland, under the projects Preludium 14 (No.~2017/27/N/ST3/01604) and Opus 11 (No.~2016/21/B/ST3/02160), by the the START scholarship of the Foundation for Polish Science (FNP) and by the Scholarships of Minister of Science and Higher Education. M. E. acknowledges funding from the European Research Council project NanoThermo (ERC-2015-CoG Agreement No. 681456) and by the National Research Fund of Luxembourg (Project No. QUTHERM C18/MS/12704391).

\end{acknowledgments}

\appendix

\section{Role of initial state} \label{sec:app}
This Appendix explores a potential role played by the initial state of the system in the post-thermalization. Specifically, Sec.~\ref{subsec:apferm} deals with noninteracting fermionic systems while Sec.~\ref{subsec:apspin} with spin chains.

\subsection{Fermionic two-level system} \label{subsec:apferm}
As shown in Sec.~\ref{subsec:irl}, interactions in fermionic systems can lead to the emergence of higher-order (non-Gaussian) correlations which may preserve the system-environment mutual information, thus suppressing the post-thermalization. One may wonder whether a similar behavior may be observed even in noninteracting systems given an initial non-Gaussian state. To be specific, initial non-Gaussianity may be present only to the system since the thermal state of a quadratic Hamiltonian of the environment is Gaussian. Furthermore, non-Gaussianity cannot be realized in the noninteracting resonant level considered in Sec.~\ref{subsec:nrl}, in which the system consists of a single fermionic level. Indeed, in that case the state of the system is always a mixture of the empty and the occupied state, which is Gaussian; this is due to the parity superselection rule which prohibits coherent superpositions of states with even and odd parity~\cite{wick1952}. Therefore, to explore the role of non-Gaussianity we consider a system consisting of a two coupled levels A and B; the whole model is described by the Hamiltonian $\hat{H}_{SE}=\hat{H}_S+\hat{H}_E+\hat{H}_I$ where
\begin{align}
	\hat{H}_S&=\epsilon_A c^\dagger_A c_A+\epsilon_B c^\dagger_B c_B + \mathcal{T} \left( c^\dagger_A c_B+c^\dagger_B c_A \right), \\
	\hat{H}_E&=\sum_k \epsilon_k c^\dagger_k c_k, \\
	\hat{H}_I&=\Omega \sum_{k} \left( c_B^\dagger c_{k} + c_{k}^\dagger c_B \right).
\end{align}
As previously, energy levels of the bath $\epsilon_k$ are uniformly distributed over the interval $[-W/2,W/2]$ and $\Omega$ is parameterized as $\Gamma=2\pi \Omega^2 (K-1)/W$. We take the initial state to be described by a random density matrix
\begin{align}
\rho_S(0)=\begin{pmatrix}
	\rho_{00,00} & 0 & 0 & \rho_{00,11}\\
0 & \rho_{01,01} & \rho_{01,10} & 0 \\
	0 & \rho_{10,01} & \rho_{10,10} & 0 \\
	\rho_{11,00} &  0 & 0 & \rho_{11,11}
\end{pmatrix}.
\end{align}
Here states of the system are denoted as $\{00,01,10,11 \}$, where first (second) digit describes the occupation of the level A (B). Random elements of the density matrix obey the conditions $\text{Tr} \rho_{S}(0)=1$, $|\rho_{00,11}|=|\rho_{11,00}| \leq \sqrt{\rho_{00,00} \rho_{11,11}}$ and $|\rho_{01,10}|=|\rho_{10,01}| \leq \sqrt{\rho_{01,01} \rho_{10,10}}$.

%
\begin{figure}
	\centering
	\includegraphics[width=0.90\linewidth]{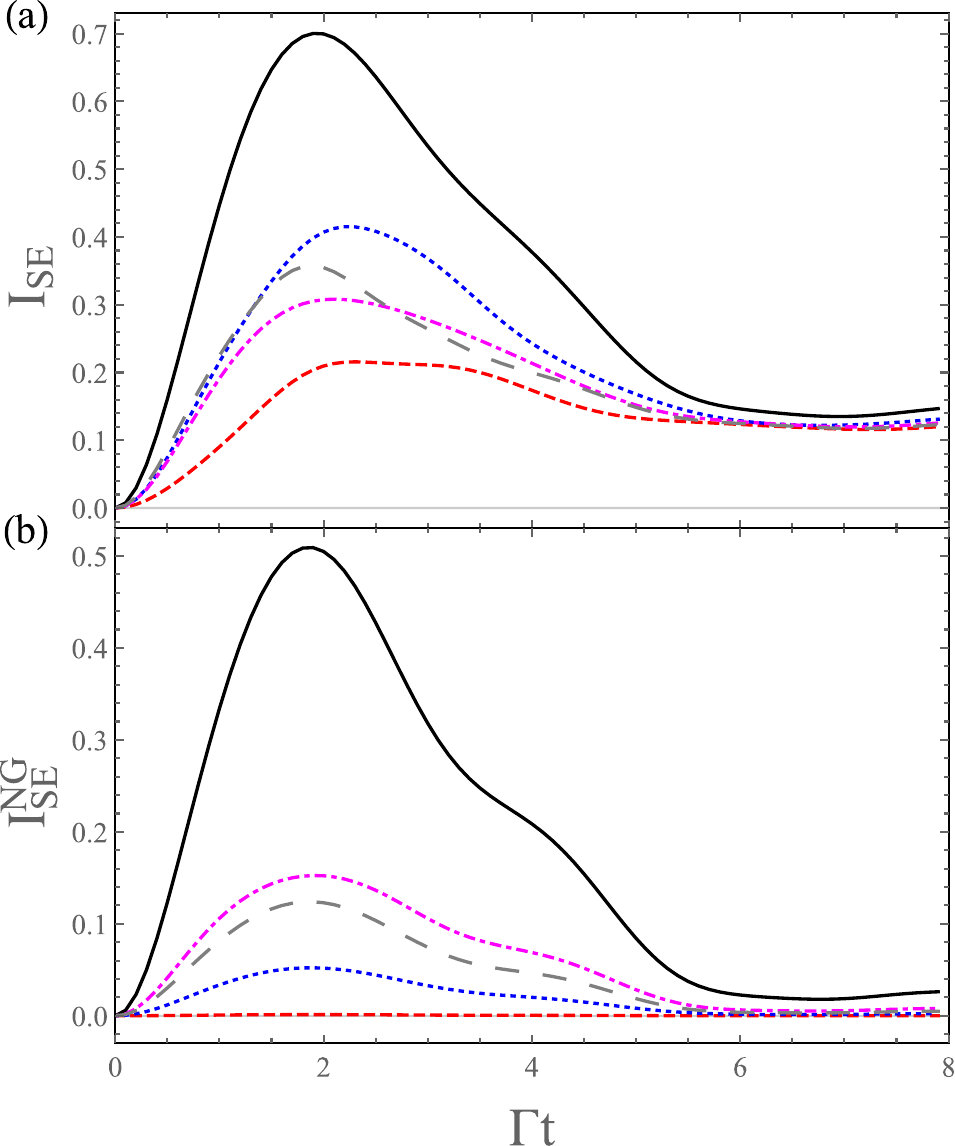}		
	\caption{The evolution of the system-environment mutual information (a) and its non-Gaussian part (b) for the two-site model with several random realizations of the initial state of the system. Parameters: $\epsilon_A=\epsilon_B=0$, $\mathcal{T}=1$, ${\Gamma=1}$, ${\mu=0}$, ${\beta=1}$, ${W=3}$ and ${K=7}$.}
	\label{fig:aferm}
\end{figure}
%
In Fig.~\ref{fig:aferm} we present a behavior of the system-environment mutual information $I_{SE}$ [Fig.~\ref{fig:aferm}~(a)] and the non-Gaussian contribution to the mutual information $I_{SE}^{NG}$ defined in Sec.~\ref{subsec:irl} [Fig.~\ref{fig:aferm}~(b)] for several randomly generated initial states of the system. It appears that the mutual information converges to an approximately the same residual value independent of the initial state. Furthermore, the non-Gaussian contribution to the mutual information is first generated, but later approaches zero, being reconverted into intra-environment non-Gaussian correlations. Therefore, the post-thermalization is observed also for non-Gaussian initial states. This suggests that interactions not only lead to generation of higher-order correlations between the system and the environment, but are also essential for their preservation. This somewhat resembles the result of Gluza \textit{et al.}~\cite{eisert2016} showing that fermionic states tend to loss their non-Gaussian features when evolving under quadratic Hamiltonians.

\subsection{Spin chain} \label{subsec:apspin}
In Sec.~\ref{subsec:censpin} we explained the lack of post-thermalization in the  central spin model using the correspondence to interacting fermionic models provided by Jordan-Wigner transformation. Nevertheless, a mapping to noninteracting fermions is applicable for a different topology of the underlying Hamiltonian, namely, when the total system-bath ensemble consists of a one-dimensional XY spin chain. This suggests that the post-thermalization may be still observed in certain spin systems. There is, however, an important difference between spin systems and fermionic ones. For a single spin, one may realize an arbitrary coherent superposition of ${|\!\!\uparrow \rangle}$ and ${|\!\!\downarrow \rangle}$ states. Jordan-Wigner transformation provides then mapping onto a coherent superposition of the occupied and the empty state, which is forbidden by the parity superselection rule. Therefore, even though a spin Hamiltonian can be always mapped onto a (either interacting or noninteracting) fermionic Hamiltonian, not all spin states correspond to physically permissible fermionic states. 

%
\begin{figure}
	\centering
	\includegraphics[width=0.90\linewidth]{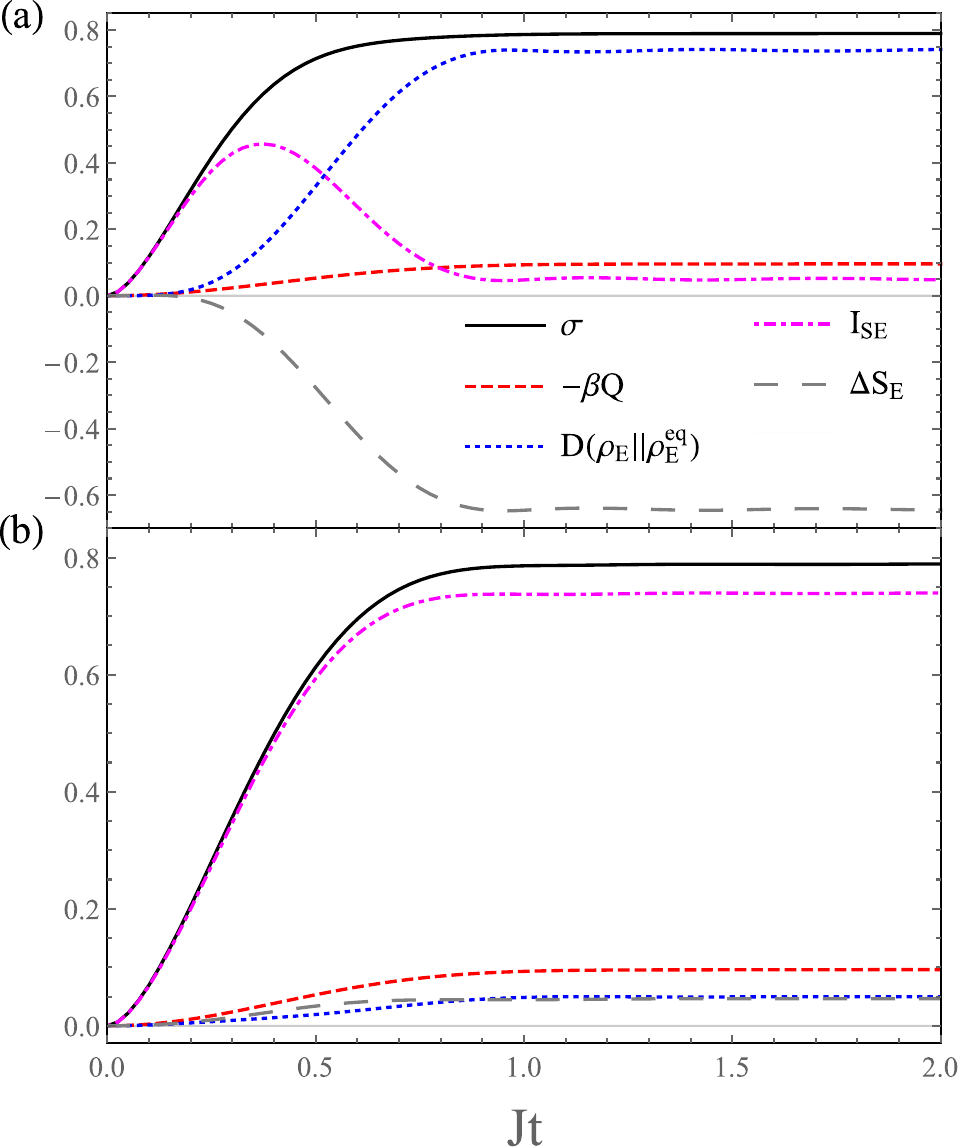}		
	\caption{The evolution of the entropy production and its constituents for the XY spin chain with the system initialized in the $|\! \! \downarrow \rangle$ state (a) and ${(|\! \!\downarrow \rangle + |\! \!\uparrow \rangle)/\sqrt{2}}$ state (b). Parameters: ${J=0.2}$, $\gamma=0.5$, ${h_d=h_E=0}$, ${\beta=1}$ and ${K=7}$.}
	\label{fig:aspin}
\end{figure}
%
Here we explore how coherence of the initial spin state may effect the post-thermalization. To this goal we consider the XY spin chain with
\begin{align}
	\hat{H}_S&=h_d \hat{s}_d^z, \\
	\hat{H}_E &= h_E \sum_{k=1}^K \hat{s}_k^z + \gamma \sum_{k=1}^{K-1} \left( \hat{s}_k^+ \hat{s}_{k+1}^-  + \hat{s}_k^- \hat{s}_{k+1}^+  \right), \\
	\hat{H}_I&= J \left(\hat{s}_d^+ \hat{s}_{1}^- +\hat{s}_d^- \hat{s}_{1}^+ \right).
\end{align}
In contrast to the central spin model, the sites in the environment are now mutually coupled and the system is attached only to a single spin of the environment. Using Jordan-Wigner transformation the model can be mapped onto a noninteracting fermionic chain with
\begin{align}
	\hat{H}_S&=h_d c_d^\dagger c_d, \\
	\hat{H}_E &= h_E \sum_{k=1}^K c_k^\dagger c_k + \gamma \sum_{k=1}^{K-1} \left( c_k^\dagger c_{k+1}+ c_{k+1}^\dagger c_k  \right), \\
	\hat{H}_I&= J \left(c_d^\dagger c_1+c_1^\dagger c_d \right).
\end{align}

Fig.~\ref{fig:aspin} presents the behavior of the entropy production and its constituents for two initial states: ${|\!\!\downarrow\rangle}$ [Fig.~\ref{fig:aspin}~(a)] and ${(|\!\!\downarrow \rangle + |\! \!\uparrow \rangle)/\sqrt{2}}$ [Fig.~\ref{fig:aspin}~(b)]. As one can observe, for the initial ${| \! \!\downarrow \rangle}$ state the system undergoes the post-thermalization, with the system-environment mutual information being reconverted into the relative entropy $D(\rho_E||\rho_E^\text{eq})$. We note that a qualitative interpretation of post-thermalization as a result of reconversion of the system-environment mutual information $I_{SE}$ into the intra-environment correlation $I_\text{env}$ does not apply to the spin chain model due to presence of initial correlations between the spins in the environment. In contrast, for the initial superposition ${(|\! \!\downarrow \rangle + |\! \!\uparrow \rangle)/\sqrt{2}}$ the system-environment mutual information remains the predominant contribution to the entropy production even at long times. Therefore, in spin models the presence of post-thermalization may depend on the initial state of the system.  While a clear explanation of this phenomenon is lacking, the preservation of $I_{SE}$ is clearly related to the presence of unphysical correlations between states with even and odd parity in the corresponding fermionic state obtained by means of Jordan-Wigner transformation.


\begin{thebibliography}{}
\bibitem{esposito2010}
M. Esposito, K. Lindenberg, and C. Van den Broeck, Entropy production as correlation between system and reservoir, \href{\doibase 10.1088/1367-2630/12/1/013013}{New J. Phys. \textbf{12}, 013013 (2010)}.

\bibitem{reeb2014}
D. Reeb and M. M. Wolf, An improved Landauer principle with finite-size corrections, \href{\doibase 10.1088/1367-2630/16/10/103011}{New J. Phys. \textbf{16}, 103011 (2014)}.

\bibitem{ptaszynski2019}
K. Ptaszy\'{n}ski and M. Esposito, Entropy Production in Open Systems: The Predominant Role of Intraenvironment
Correlations, \href{\doibase 10.1103/PhysRevLett.123.200603}{Phys. Rev. Lett. \textbf{123}, 200603 (2019)}.

\bibitem{araki1970}
H. Araki and E. H. Lieb, Entropy inequalities, \href{\doibase 10.1007/BF01646092}{Commun. Math. Phys. \textbf{18}, 160 (1970)}.

\bibitem{jaeger2007}
G. Jaeger, \textit{Quantum Information: An Overview} (Springer, New York, 2007).

\bibitem{pucci2013}
L. Pucci, M. Esposito, and L. Peliti, Entropy production in quantum Brownian motion, \href{\doibase 10.1088/1742-5468/2013/04/P04005}{J. Stat. Mech. (2013) P04005}.

\bibitem{einsiedler2020}
S. Einsiedler, Non-perturbative approach to non-Markovianity and entropy production in the Caldeira-Leggett model, \href{\doibase 10.6094/UNIFR/165942}{Master's thesis, Albert-Ludwigs-Universit\"{a}t Freiburg, 2020}.

\bibitem{cusumano2018}
S. Cusumano, V. Cavina, M. Keck, A. De Pasquale, and V. Giovannetti, Entropy production and asymptotic factorization via thermalization: a collisional model approach, \href{\doibase 10.1103/PhysRevA.98.032119}{Phys. Rev. A \textbf{98}, 032119 (2018)}.

\bibitem{nandkishore2015}
R. Nandkishore and D. A. Huse, Many-Body Localization and Thermalization in Quantum Statistical Mechanics, \href{\doibase 10.1146/annurev-conmatphys-031214-014726}{Ann. Rev. Cond. Mat. Phys. \textbf{6}, 15 (2015)}.

\bibitem{bohrdt2017}
A. Bohrdt, C. B. Mendl, M. Endres, and M. Knap, Scrambling and thermalization in a diffusive quantum many-body system, \href{\doibase 10.1088/1367-2630/aa719b}{New J. Phys \textbf{19}, 063001 (2017)}.

\bibitem{swingle2018}
B. Swingle, Unscrambling the physics of out-of-time-order correlators, \href{\doibase 10.1038/s41567-018-0295-5}{Nat. Phys. \textbf{14}, 988 (2018)}.

\bibitem{swan2019} 
R. J. Lewis-Swan, A. Safavi-Naini, J. J. Bollinger, and A. M. Rey, Unifying scrambling, thermalization and entanglement through measurement of fidelity out-of-time-order correlators in the Dicke model, \href{\doibase 10.1038/s41467-019-09436-y}{Nat. Commun. \textbf{10}, 1581 (2019)}.

\bibitem{hummel2019} 
Q. Hummel, B. Geiger, J. D. Urbina, and K. Richter, Reversible Quantum Information Spreading in Many-Body Systems near Criticality, \href{\doibase 10.1103/PhysRevLett.123.160401}{Phys. Rev. Lett. \textbf{123}, 160401 (2019)}.

\bibitem{peschel2003}
I. Peschel, Calculation of reduced density matrices from correlation functions, \href{\doibase 10.1088/0305-4470/36/14/101}{J. Phys. A: Math. Gen. \textbf{36}, L205 (2003)}.

\bibitem{eisler2012}
V. Eisler and I. Peschel, On entanglement evolution across defects in critical chains, \href{\doibase 10.1209/0295-5075/99/20001}{Europhys. Lett. \textbf{99}, 20001 (2012)}.

\bibitem{sharma2015}
A. Sharma and E. Rabani, Landauer current and mutual information, \href{\doibase 10.1103/PhysRevB.91.085121}{Phys. Rev. B \textbf{91}, 085121 (2015)}.

\bibitem{colla2021}
A. Colla and H.-P. Breuer, Entropy production and the role of correlations in quantum Brownian motion, \href{\doibase 10.1103/PhysRevA.104.052408}{Phys. Rev. A \textbf{104}, 052408 (2021)}.

\bibitem{cai2014}
C. Y. Cai, L.-P. Yang, and C. P. Sun, Threshold for nonthermal stabilization of open quantum systems, \href{\doibase 10.1103/PhysRevA.89.012128}{Phys. Rev. A \textbf{89}, 012128 (2014)}.

\bibitem{xiong2015}
H.-N. Xiong, P.-Y. Lo, W.-M. Zhang, D. H. Feng, and F. Nori, Non-Markovian Complexity in the Quantum-to-Classical Transition, \href{\doibase 10.1038/srep13353}{Sci. Rep. \textbf{5}, 13353 (2015)}.

\bibitem{jussiau2019}
\'{E}. Jussiau, M. Hasegawa, and R. S. Whitney, Signature of the transition to a bound state in thermoelectric quantum transport, \href{\doibase 10.1103/PhysRevB.100.115411}{Phys. Rev. B \textbf{100}, 115411 (2019)}.

\bibitem{alvaredo2020}
O. A. Castro-Alvaredo, M. Lencs\'{e}s, I. M. Sz\'{e}cs\'{e}nyi, and J. Viti, Entanglement Oscillations near a Quantum Critical Point, \href{\doibase 10.1103/PhysRevLett.124.230601}{Phys. Rev. Lett. \textbf{124}, 230601 (2020)}.

\bibitem{nghiem2016}
H. T. M. Nghiem, D. M. Kennes, C. Kl\"{o}ckner, V. Meden, and T. A. Costi, Ohmic two-state system from the perspective of the interacting resonant level model: Thermodynamics and transient dynamics, \href{\doibase 10.1103/PhysRevB.93.165130}{Phys. Rev. B \textbf{93}, 165130 (2016)}.

\bibitem{genoni2008}
M. G. Genoni, M. G. A. Paris, and K. Banaszek, Quantifying the non-Gaussian character of a quantum state by quantum relative entropy, \href{\doibase 10.1103/PhysRevA.78.060303}{Phys. Rev. A \textbf{78}, 060303(R) (2008)}.

\bibitem{backens2019}
S. Backens, A. Shnirman, and Y. Makhlin, Jordan--Wigner transformations for tree structures, \href{\doibase 10.1038/s41598-018-38128-8}{Sci. Rep. \textbf{9}, 2598 (2019)}.

\bibitem{gring2012}
M. Gring, M. Kuhnert, T. Langen, T. Kitagawa, B. Rauer, M. Schreitl, I. Mazets, D. A. Smith, E. Demler, and J. Schmiedmayer, Relaxation and Prethermalization in an Isolated Quantum System, \href{\doibase 10.1126/science.1224953}{Science \textbf{337}, 1318 (2012)}.

\bibitem{mori2018}
T. Mori, T. N. Ikeda, E. Kaminishi, and M. Ueda, Thermalization and prethermalization in isolated quantum systems: a theoretical overview, \href{\doibase 10.1088/1361-6455/aabcdf}{J. Phys. B \textbf{51}, 112001 (2018)}.

\bibitem{liu2014}
H. Liu and S. J. Suh, Entanglement growth during thermalization in holographic systems, \href{\doibase 10.1103/PhysRevD.89.066012}{Phys. Rev. D \textbf{89}, 066012 (2014)}.

\bibitem{breuer2002}
H.-P. Breuer and F. Petruccione, \textit{The Theory of Open Quantum Systems} (Oxford University Press, Oxford, 2002).

\bibitem{wick1952}
G. C. Wick, A. S. Wightman, and E. P. Wigner, The Intrinsic Parity of Elementary Particles, \href{\doibase 10.1103/PhysRev.88.101}{Phys. Rev. \textbf{88}, 101 (1952)}.

\bibitem{eisert2016}
M. Gluza, C. Krumnow, M. Friesdorf, C. Gogolin, and J. Eisert, Equilibration via Gaussification in Fermionic Lattice Systems, \href{\doibase 10.1103/PhysRevLett.117.190602}{Phys. Rev. Lett. \textbf{117}, 190602 (2016)}.

\end{thebibliography}
\end{document}